Anomalous conduction band fluctuation in silicon carbide metal-oxide-semiconductor structures revealed by electrical characterization of field effect transistors combined with self-consistent numerical calculations

Koji Ito[1], Takuma Kobayashi[1], and Tsunenobu Kimoto[1]

[1]Department of Electronic Science and Engineering, Kyoto University, Nishikyo, Kyoto 615-8510, Japan

We determined the interface state density ($D_{it}$) distributions in the vicinity of the conduction band edge in silicon carbide (SiC) metal-oxide-semiconductor (MOS) structures by reproducing the experimental current-voltage characteristics of MOS field effect transistors (MOSFETs) with numerical calculations. In the calculation, potential distributions and energy sub-bands in the inversion layer were calculated by solving Poisson and Schrödinger equations, respectively. We demonstrate that gate characteristics of the MOSFETs are well described by considering that the interface states are caused by fluctuation of free-electron density of states in a two-dimensional system. The $D_{it}$ distributions are almost uniquely determined by the oxide formation process (as oxidation or interface nitridation) and independent of the acceptor concentration ($3\times10^{15} - 1\times10^{18}$ cm$^{-3}$).

The interface between a semiconductor and its oxide is a stage of transistor action in electron devices and its characterization is essential in nanoscience and technology. An example which has been studied intensively is the interface of silicon carbide (SiC) and silicon dioxide ($SiO_2$). SiC is a compound semiconductor owning unique physical properties for power device applications, such as a wide band gap and high breakdown electric field [1, 2]. Thus, SiC metal-oxide-semiconductor field effect transistors (MOSFETs) have been attracting increasing attention as low-loss and high-frequency power switching devices. SiC MOSFETs have, however, suffered from the low channel mobility due to the high interface state density ($D_{it}$) in $SiO_2$/SiC structures ($\sim 10^{13}$ cm$^{-2}$eV$^{-1}$) [3–14]. Although it was found that interface nitridation (annealing in nitric oxide (NO) [15–18] or nitrous oxide ($N_2O$) [15, 19, 20]) or phosphorus treatment (annealing in gas mixture of phosphoryl chloride ($POCl_3$), oxygen ($O_2$) and, nitrogen ($N_2$) [21, 22]) can reduce the $D_{it}$ and improve the channel mobility to some extent, the physical origin of the abnormally high interface state density is still not understood.

In recent years, SiC has also been regarded as an attractive material in the field of quantum computing. Some of the bulk defects in SiC, such as silicon vacancies ($V_{Si}$) [23], divacancies ($V_{Si}$-$V_C$) [24], and antisite-vacancy pairs ($C_{Si}$-$V_C$) [25] were suggested to be promising for qubits (i.e., single photon sources), similar to the well-known nitrogen-vacancy (NV) center in diamond [26]. Interface defects also play an important role in this field; Single-photon-emitting SiC diodes operating up to room temperature were reported [27], that might originate from interface (carbon) defects [28]. Thus, understanding the physical origin of interface states in SiC is highly demanded in the fields of both power electronics and quantum computing.

In previous studies, it was indicated that the $D_{it}$ values rapidly increase with approaching the conduction band edge ($E_C$) [9, 29]. Thus, determination of the $D_{it}$ in the shallow energy range near $E_C$ is particularly important in understanding the nature of the interface states. However, few reports have focused on the $D_{it}$ distributions in the vicinity of $E_C$ so far, since they cannot be simply estimated from

capacitance–voltage ($C$–$V$) characteristics of MOS capacitors [30]; A study using MOS Hall effect and split $C$–$V$ measurements indicated that the $D_{it}$ values very near $E_C$ are reduced with the interface nitridation [9]. A more recent study has suggested the possibility that the interface states originate from the *tail* of two-dimensional density of states (2D-DOS) of SiC by characterizing wet-oxidized MOSFETs at cryogenic temperatures [31]. In order to understand the nature of interface states, however, a more systematic study involving MOS structures with various acceptor densities and oxide formation conditions is demanded.

In this study, we determined the $D_{it}$ distributions for SiC MOS structures in the vicinity of $E_C$ by reproducing the experimental gate characteristics of MOSFETs with numerical calculations. In the calculation, potential distributions and energy sub-bands in the inversion layer were calculated by solving Poisson and Schrödinger equations, respectively. We prepared MOSFETs with various acceptor densities and oxidation formation conditions to discuss their impacts on the interface properties. The limiting factors of drain current in SiC MOSFETs are also discussed based on the results.

MOSFETs were fabricated on 8° off-axis p-type (acceptor density: $N_A = 3 \times 10^{15} - 1 \times 10^{18}$ cm$^{-3}$) 4H-SiC (0001) epilayers. The gate oxides were formed by dry oxidation at 1300°C for 30 min (As-Ox.) or by dry oxidation with subsequent annealing in NO (10% diluted in nitrogen (N$_2$)) at 1250°C for 70 min (Ox.+NO), resulting in an oxide thickness of about 42 nm. The p-type body regions of MOSFETs were formed either by epitaxial growth ($N_A = 3 \times 10^{15}$ cm$^{-3}$) or by aluminum (Al) implantation ($N_A = 3 \times 10^{16} - 1 \times 10^{18}$ cm$^{-3}$). The source/drain regions were formed by high-dose phosphorus (P) implantation ($N_D = 1 \times 10^{20}$ cm$^{-3}$) to obtain good ohmic contacts. The channel length and width of the MOSFETs were 50 or 100 μm and 200 μm, respectively. All of the measurements were conducted at room temperature (RT).

Figure 1 shows a schematic illustration describing the concept of the model to extract the $D_{it}$

distribution from gate characteristics of MOSFETs [32]. In this model, it is assumed that the electrons in the inversion layer contribute to the conduction with constant drift mobility, and that the electrons trapped at the interface states are completely immobile. In general, the drain current ($I_D$) and the gate voltage ($V_G$) are expressed as a function of the surface potential ($\psi_S$) by [33]

$$I_D(\psi_S) = \frac{W}{L} e n_{\text{free}}(\psi_S) \mu_{\text{drift}}(\psi_S) V_D, \quad (1)$$

$$V_G(\psi_S) = V_{\text{FB}} + \psi_S + \frac{Q_{\text{fix}} + Q_{\text{dep}}(\psi_S) + e n_{\text{free}}(\psi_S) + e n_{\text{trap}}(\psi_S)}{C_{\text{ox}}}. \quad (2)$$

Here, $W$ is the channel width, $L$ the channel length, $e$ the elementary charge, $\mu_{\text{drift}}$ the drift mobility, $n_{\text{free}}$ the density of free electrons in the channel, $n_{\text{trap}}$ that of trapped electrons, $V_D$ the drain voltage, $V_{\text{FB}}$ the flat band voltage, $Q_{\text{fix}}$ the fixed charge, $Q_{\text{dep}}$ the depletion charge, and $C_{\text{ox}}$ the oxide capacitance. $Q_{\text{dep}}$ is further approximated by [33]

$$Q_{\text{dep}}(\psi_S) = \sqrt{2e\varepsilon_s N_A \psi_S}, \quad (3)$$

where $\varepsilon_S$ is the dielectric constant of the semiconductor.

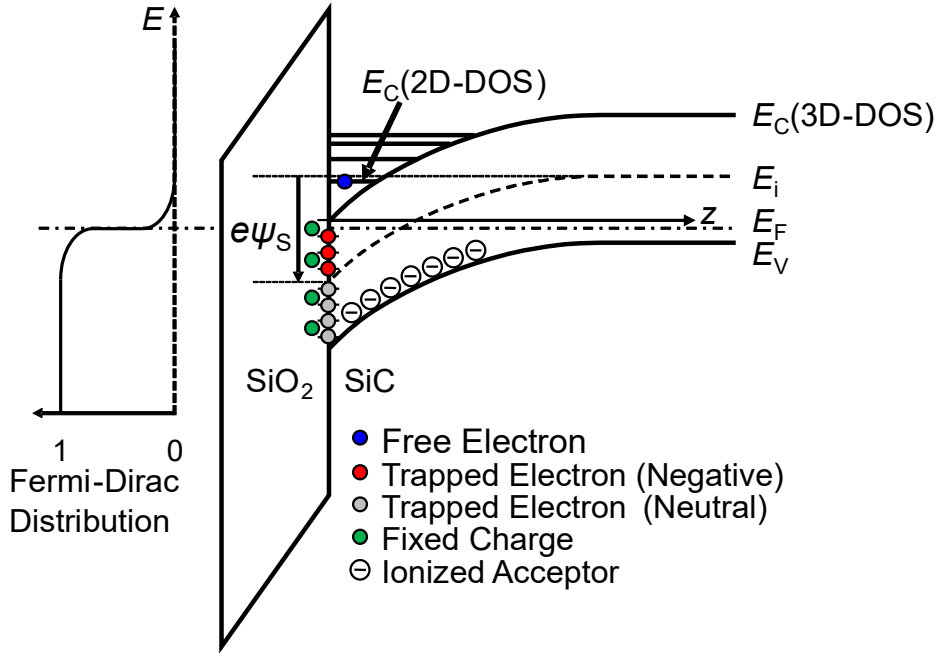

Fig. 1: Concept of a calculation model for extracting interface state density ($D_{it}$) distributions from gate characteristics of SiC MOSFETs. $E_C$, $E_i$, $E_F$, $E_V$, and $\psi_S$ indicate the conduction band edge, the intrinsic level, the Fermi level, the valence band edge, and the surface potential of SiC, respectively.

In using the model above, we assumed constant $\mu_{\text{drift}}$ of 100, 35, 25, 15 and 5 cm$^2$V$^{-1}$s$^{-1}$ for MOSFETs with $N_A$ of 3×10$^{15}$, 3×10$^{16}$, 1×10$^{17}$, 3×10$^{17}$ and 1×10$^{18}$ cm$^{-3}$, respectively, regardless of oxide formation conditions (As-Ox. or Ox.+NO), based on the reported values of Hall mobility for NO-annealed MOSFETs [34]. Indeed, Ref. [9] indicated that the difference in the Hall mobility between as-oxidized and NO-annealed MOSFETs is considerably small (e.g. about 160 and 140 – 180 cm$^2$V$^{-1}$s$^{-1}$ at $n_{\text{free}}$ = 10$^{11}$ cm$^{-2}$ for as-oxidized and NO annealed MOSFETs, respectively), making such assumption reasonable. As a result of MOS Hall effect measurement, we confirmed that $\mu_{\text{drift}}$ was in the range of 83 – 121 cm$^2$V$^{-1}$s$^{-1}$ for a lightly-doped ($N_A$ = 3×10$^{15}$ cm$^{-3}$) MOSFET, which also guarantee the validity of the assumed mobility.

By numerically solving the Schrödinger and Poisson equations with setting $n_{\text{free}}$ and $N_A$ as initial values such as 1×10$^{12}$ cm$^{-2}$ and 3×10$^{15}$ cm$^{-3}$, respectively, $E_F$, $\psi_S$ and the energy level of the first sub-band of 2D-DOS were determined [35, 36]. Then, by assuming a triangle-shaped potential ($\varphi(z)$) at the MOS interface we obtained the initial function of $\varphi(z)$. As the next step, we solved the Schrödinger equation with the effective mass approximation, that is [35]

$$\left[-\frac{\hbar^2}{2m^*_{c_\parallel}}\frac{d^2}{dz^2} - e\phi(z)\right]\xi_i(z) = E_i\xi_i(z), \qquad (4)$$

where $\hbar$ is the Dirac's constant, $m^*_{c_\parallel}$ the effective mass parallel to the $c$-axis, $i$ the number of the sub-band of energy eigenvalues, $\xi_i(z)$ the wave function, and $E_i$ the energy eigenvalues. As a result of this calculation, $\xi_i(z)$ and the $E_i$ were obtained. We then determined $E_F$ and the electron density at each sub-band in the inversion layer ($n_i$) so as to satisfy the following equations [35, 37]:

$$D_{2D} = M_C \frac{m^*_{c_\perp}}{\pi\hbar^2}, \qquad (5)$$

$$n_i = D_{2D}k_BT \ln\left[1 + \exp\left(\frac{E_F - E_i}{k_BT}\right)\right], \qquad (6)$$

$$n_{\text{free}} = \sum_{i} n_i, \tag{7}$$

where $M_C$ (= 3 [2]) is the number of equivalent valleys, $m_{c_\perp}^*$ the effective mass perpendicular to the $c$-axis, $k_B$ the Boltzmann constant, and $T$ the absolute temperature. A sufficient number of the sub-bands was taken into consideration up to the energy level at which the occupied electron density was negligible. The lowest and second lowest energy bands with different effective masses are considered; Table I summarizes the effective masses of those bands [38]. The band edge of the second lowest energy band is located at approximately 0.12 eV higher than that of the lowest band.

The Poisson equation was solved based on the obtained $n_i$ and the $\xi_i$ above to determine $\varphi(z)$ for the next step [35]:

$$\frac{d^2 \varphi(z)}{dz^2} = -\frac{\rho_{\text{dep}}(z) - e \sum_i n_i |\xi_i(z)|^2}{\varepsilon_s}, \tag{8}$$

where $\rho_{\text{dep}}(z)$ (= $-eN_A$) is the charge density inside the depletion layer. The above calculation procedure was repeated until the difference of $E_F$ between $N$ and $(N-1)$ steps ($N$: natural number) was less than $10^{-6}$ eV.

Table I: Effective masses for the lowest and the second lowest energy bands in 4H-SiC [38]. Here, $m_e$ is the electron rest mass. In calculating the effective masses, we applied $m_{c_\perp}^* = \sqrt{m_{M\Gamma} m_{MK}}$ and $m_{c_\parallel}^* = m_{ML}$.

|  | Lowest Band | Second Lowest Band |
|---|---|---|
| $m_{ML}$ | $0.31 m_e$ | $0.71 m_e$ |
| $m_{M\Gamma}$ | $0.57 m_e$ | $0.78 m_e$ |
| $m_{MK}$ | $0.28 m_e$ | $0.16 m_e$ |
| $m_{c_\perp}^* = \sqrt{m_{M\Gamma} m_{MK}}$ | $0.40 m_e$ | $0.35 m_e$ |
| $m_{c_\parallel}^* = m_{ML}$ | $0.31 m_e$ | $0.71 m_e$ |

Finally, $n_{trap}$ was calculated by [39]

$$n_{trap}(\psi_S) = \int_{E_i}^{\infty} \frac{D_{it}(\psi_S)}{1 + \exp\left(\frac{E - E_F}{k_B T}\right)} dE. \qquad (9)$$

Here, $D_{it}$ distribution was assumed to be expressed as [6, 40]

$$D_{it} = D_0 + D_1 \exp\left(\frac{E - E_C}{E_1}\right) + D_2 \exp\left(\frac{E - E_C}{E_2}\right), \qquad (10)$$

where $D_0$, $D_1$, $D_2$, $E_1$, $E_2$, and $Q_{fix}$ were used as fitting parameters to reproduce the experimental gate characteristics.

Figure 2 shows the typical experimental and calculated gate characteristics for a NO-annealed MOSFET ($N_A = 3 \times 10^{15}$ cm$^{-3}$, peak field-effect mobility: 31 cm$^2$V$^{-1}$s$^{-1}$). The calculated result well reproduces the experimental characteristic in the range of $V_G = 0 - 15$ V (corresponding to energy range of $E_C - E_T = -0.04 - 0.19$ eV).

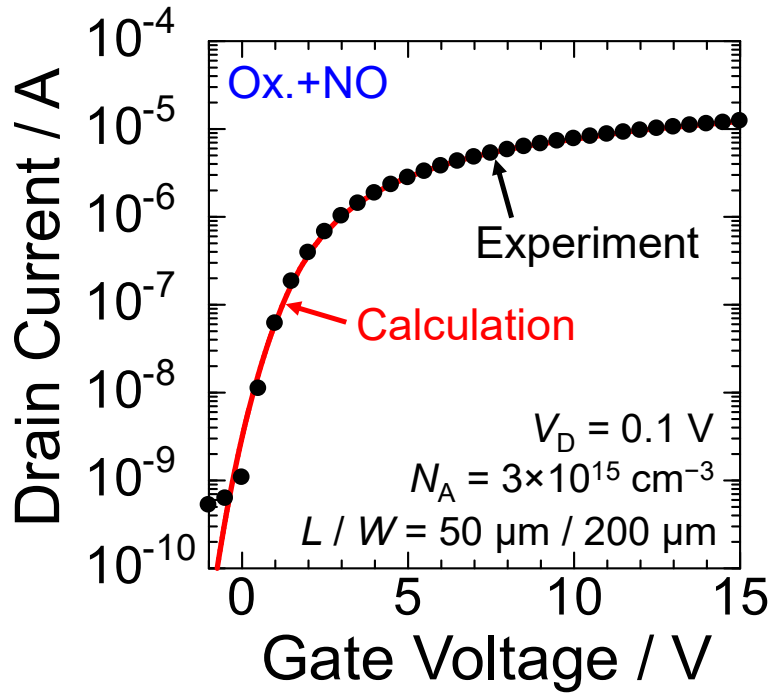

Fig. 2: Typical experimental and calculated gate characteristics for a lightly-doped NO-annealed MOSFET ($N_A = 3 \times 10^{15}$ cm$^{-3}$, peak field-effect mobility: 31 cm$^2$V$^{-1}$s$^{-1}$).

Figure 3(a) shows the $D_{it}$ distributions plotted with respect to the bottom edge of 3D-DOS obtained from the $I_D$–$V_G$ fitting. The obtained $D_{it}$ distributions strongly depend on the acceptor concentration of p-body. On the other hand, Figure 3(b) shows the $D_{it}$ distributions plotted with respect to the bottom edge of 2D-DOS. In contrast to the results based on the bottom edge of 3D-DOS (Fig.3(a)), the $D_{it}$ distributions are almost uniquely determined by the oxide formation condition (As-Ox. or Ox.+NO) and independent of acceptor concentration. This result suggests that the interface states near the conduction band edge in SiC MOS systems are caused by the energy fluctuation of 2D-DOS.

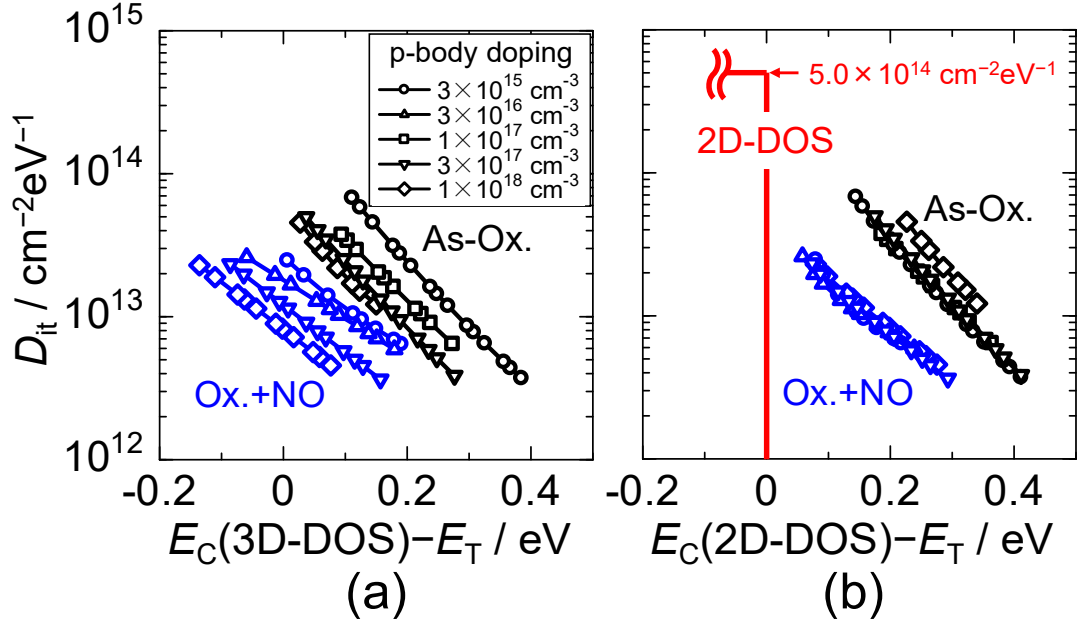

Fig. 3: Energy distributions of interface state density with respective to the bottom edge of (a) three-dimensional and (b) two-dimensional density of states obtained from the gate characteristics of MOSFETs.

Based on the obtained results, we briefly discuss the dominant limiting factors of the drain current in the SiC MOSFETs. Figure 4 shows the ratio of free electron density ($n_{free}$) to the total electron density ($n_{free} + n_{trap}$) for the fabricated MOSFETs as a function of the free electron density dependence.

In Fig. 4, we see that the $n_{\text{free}}/(n_{\text{free}} + n_{\text{trap}})$ ratio is about two to four times higher in the NO-annealed samples than in the as-oxidized samples at a given $n_{\text{free}}$, indicating that the drain current increase by the NO annealing is attributable to the increase in the free electron density. The $n_{\text{free}}/(n_{\text{free}} + n_{\text{trap}})$ ratio is almost identical among samples with different acceptor densities. On the other hand, it is known that the drain current remarkably decreases in the heavily-doped MOSFETs [14, 41]. Thus, the drain current decrease in the heavily-doped MOSFETs is mainly due to the decrease in the drift mobility, rather than the decrease in the free electron density.

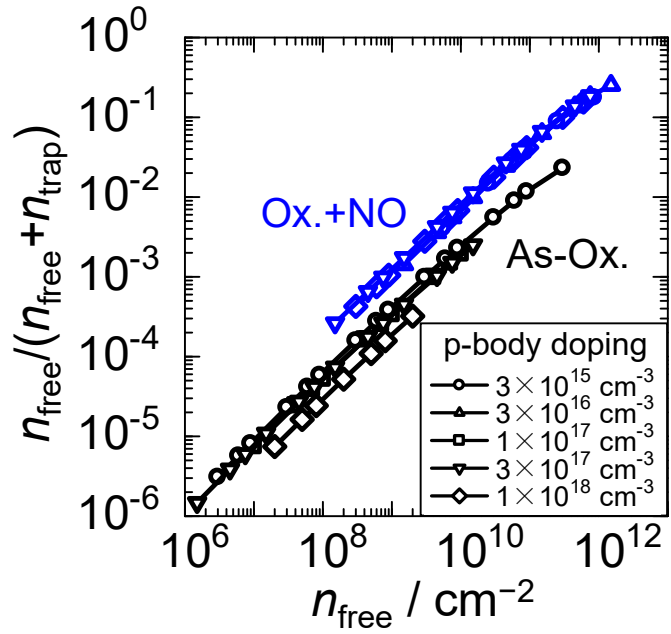

Fig. 4: Proportion of free electrons to the total electrons (= $n_{\text{free}}/(n_{\text{free}} + n_{\text{trap}})$) as a function of the free electron density for SiC MOSFETs with various acceptor concentrations of p-bodies.

In conclusion, we determined the $D_{\text{it}}$ distributions near the conduction band edge in SiC MOS structures by reproducing the experimental $I_{\text{D}}$–$V_{\text{G}}$ characteristics of MOSFETs with numerical calculations. In the calculation, potential distributions and energy sub-bands in the inversion layer were calculated by solving Poisson and Schrödinger equations, respectively. The obtained $D_{\text{it}}$ distributions are almost uniquely determined by the oxide formation process (as oxidation or interface nitridation) and independent of the acceptor concentration ($3\times10^{15}$ – $1\times10^{18}$ cm$^{-3}$). We demonstrate that gate characteristics of the MOSFETs are well described by considering that the interface states are caused by the fluctuation of free-electron density of states in a two-dimensional system. We confirmed that the ratio of the free electron density with respect to the total electron density increases by the NO annealing, suggesting that the drain current increase in the NO-annealed MOSFETs is attributable to the increase in the free carrier density. In contrast, the $n_{\text{free}}/(n_{\text{free}} + n_{\text{trap}})$ is almost identical among MOSFETs with different acceptor concentrations ($3\times10^{15}$ – $1\times10^{18}$ cm$^{-3}$), indicating that the drain current decrease observed in heavily-doped MOSFETs is mainly ascribed to the decrease in the drift mobility rather than the decrease in the free carrier density.

We thank Dr. M. Horita and Prof. J. Suda at Nagoya University for providing the Hall effect measurement system. This work was supported in part by the Super Cluster Program and Open Innovation Platform with Enterprises, Research Institute and Academia (OPERA) Program from the Japan Science and Technology Agency.